\documentclass[a4paper,11pt]{article}
\pdfoutput=1 

\usepackage{jheppub} 

\usepackage[T1]{fontenc} 
\usepackage{mathtools}

\usepackage{enumitem}

\usepackage{aas_macros}
\usepackage{amsmath,amssymb}
\usepackage{multirow}
\usepackage{cleveref}
\usepackage{siunitx}
\usepackage{enumerate}
\usepackage[dvipsnames]{xcolor}
\usepackage{float}

\usepackage[compat=1.1.0]{tikz-feynman}
\usepackage{afterpage}

\usepackage{amsthm}

\usepackage{psfrag}
\usepackage{color}
\usepackage{graphicx}
\usepackage{upgreek}
\usepackage{feynmp}
\DeclareMathAlphabet{\mathpzc}{OT1}{pzc}{m}{it}

\usepackage[mathscr]{euscript}

\usepackage[toc,page]{appendix}
\usepackage{circledsteps}
\usepackage{natbib}
\setcitestyle{square,comma,numbers,sort&compress}
\usepackage{enumerate}
\usepackage{tabularx,booktabs}
\usepackage[dvipsnames]{xcolor}
\usepackage{comment}
\usepackage[caption=false]{subfig}
\usepackage[compat=1.1.0]{tikz-feynman} 
\def\mc#1{\mathcal#1}
\usepackage[section]{placeins}
\usepackage{cleveref}

\allowdisplaybreaks[4]

\definecolor{darkgreen}{rgb}{0,0.5,0}

\newcommand{\beq}{\begin{eqnarray}}
\newcommand{\eeq}{\end{eqnarray}}

\usepackage[normalem]{ulem}

\newcommand{\bseq}{\begin{subequations}}
\newcommand{\eseq}{\end{subequations}}
\newcommand{\be}{\begin{equation}}
\newcommand{\ee}{\end{equation}}

\newcommand{\blue}{\color{black}}

\newcommand{\beqa}{\begin{eqnarray}}
\newcommand{\eeqa}{\end{eqnarray}}

\usepackage{simpler-wick}
\newcolumntype{Y}{>{\centering\arraybackslash}X}

\def\mc#1{\mathcal#1}
\title{Towards distinguishing Dirac from Majorana neutrino mass with gravitational waves}

\author[a]{Stephen F. King,}
\author[b]{Danny Marfatia,}
\author[a, c]{and Moinul Hossain Rahat}

\affiliation[a]{
School of Physics and Astronomy, University of Southampton,\\
Southampton SO17 1BJ, United Kingdom }
\affiliation[b]{Department of Physics and Astronomy, University of Hawaii at Manoa, \\
Honolulu, HI 96822, USA}
\affiliation[c]{Instituto de F\'isica Corpuscular, Universidad de Valencia and CSIC,\\ Edificio Institutos Investigaci\'on, Catedr\'atico
Jos\'e Beltr\'an 2, 46980 Paterna, Spain}
\emailAdd{s.f.king@soton.ac.uk}
\emailAdd{dmarf8@hawaii.edu}
\emailAdd{moinul.rahat@ific.uv.es}

\abstract{
We propose a new method towards distinguishing the Dirac versus Majorana nature of neutrino masses from the spectrum of gravitational waves (GWs) associated with neutrino mass genesis.
Motivated by the principle of generating small neutrino masses without tiny Yukawa couplings,
we assume generic seesaw mechanisms for both
Majorana and Dirac neutrino masses. For Majorana neutrinos, we further assume a spontaneously broken gauged 
$U(1)_{B-L}$ symmetry, independently of the type of Majorana seesaw mechanism, which gives a cosmic string induced GW signal flat over a wide range of frequencies. For Dirac neutrinos, we assume the spontaneous breaking of a $Z_2$ symmetry, the minimal symmetry choice associated with all Dirac seesaw mechanisms, which is softly broken, generating a peaked GW spectrum from the annihilation of the resulting domain walls. 
In fact, the GW spectra for all types of Dirac seesaws with such a broken $Z_2$ symmetry are identical, subject to a mild caveat.
As an illustrative example, we study the simplest respective type-I seesaw mechanisms, and show that 
the striking difference in the shapes of the GW spectra can help differentiate between these Dirac and Majorana seesaws, complementing results of neutrinoless double beta decay experiments. We also discuss detailed implications of the recent NANOGrav data for Majorana and Dirac seesaw models.
}

\begin{document}
\maketitle
\flushbottom

\section{Introduction}

Charged fermion masses in the standard model (SM) are necessarily of the Dirac type because of electric charge conservation.
Neutrino mass, on the other hand, may be of two types: Dirac or Majorana, where the latter possibility arises due to the fact that neutrinos are electrically neutral.
If neutrinos were massless particles, as originally envisioned in the SM, their nature, i.e. Dirac vs  Majorana, would not be distinguishable in weak interactions. However, as oscillation experiments have confirmed, they possess nonzero, albeit tiny, 
mass~\cite{Super-Kamiokande:1998kpq, SNO:2002tuh, KamLAND:2002uet}.

Dirac and Majorana neutrino mass are traditionally distinguished experimentally by neutrinoless double beta decay \cite{Schechter:1981bd,  DellOro:2016tmg}. This process is allowed only in the former case, where the first entry of the Majorana neutrino mass matrix $m_{\beta \beta}$ is model dependent. Extensive experimental efforts are currently underway for detecting neutrinoless double beta decay, achieving upper bounds $|m_{\beta \beta}| \lesssim \mc O(10-100)$ meV \cite{KamLAND-Zen:2016pfg}, and are expected to gain another order of sensitivity in the next decade \cite{Barabash:2019drl}. However, the non-observation of neutrinoless double beta decay will not be decisive about the Majorana or Dirac nature of neutrinos. At this point, it remains interesting to seek astrophysical or cosmological probes of distinguishing the nature of the neutrino mass (see \cite{Abazajian:2019oqj,Lunardini:2019zob,Adshead:2020ekg, Hernandez-Molinero:2022zoo} for recent studies), and this work is motivated by such considerations.

We begin by recalling that 
neutrino mass is associated with the breaking of separate lepton numbers $L_e,L_{\mu},L_{\tau}$. Dirac neutrino mass preserves total lepton number $L=L_e+L_{\mu}+L_{\tau}$, while Majorana mass breaks it. In the latter case, the small mass of the neutrinos may originate from the dimension-five Weinberg operator $\bar{\ell} \ell H H$, where $\ell$ represents the lepton doublets and $H$ is a Higgs doublet, breaking the lepton number by two units. This could be associated with the {spontaneous} breaking of an Abelian $U(1)_L$ symmetry which may be global, or, when combined with baryon number, gauged $U(1)_{B-L}$. The occurrence of cosmic strings in the early universe is a consequence in both scenarios, and their subsequent decay can produce detectable gravitational wave (GW) signatures \cite{Auclair:2019wcv, Dror:2019syi, Blasi:2020wpy}. This offers a well known possible observable indication of Majorana neutrino mass.

A convincing ultraviolet completion of the Weinberg operator is achieved by introducing right-handed neutrinos that get large Majorana masses after spontaneous breaking of a $U(1)_L$ or $U(1)_{B-L}$ symmetry, the latter case opening up the possibility of a gauged Abelian symmetry which may be anomaly free if there are precisely three right-handed neutrinos.
The type-I seesaw mechanism \cite{Minkowski:1977sc,Yanagida:1979as,GellMann1979,Glashow:1979nm,Mohapatra:1979ia} then provides an elegant explanation for the generation of light neutrino masses, avoiding the need for extremely small Yukawa couplings. {Other realizations of the Weinberg operator, such as the type-II \cite{Magg:1980ut, Cheng:1980qt, Lazarides:1980nt, Mohapatra:1980yp, Schechter:1980gr} and type-III \cite{Foot:1988aq} seesaw mechanisms that involve different intermediate particles, can also generate small neutrino masses without tiny Yukawa couplings. All these mechanisms may yield cosmic string induced GW signals if the $U(1)_{L}$ or $U(1)_{B-L}$ symmetry is spontaneously broken. This provides a generic signature for all seesaw mechanisms responsible for Majorana masses.}

If neutrinos are Dirac particles, the most minimal extension of the SM would be to add two or three right-handed SM singlet neutrinos $\nu_R$ with tiny 
tree-level Yukawa couplings $y_D$, defined by $y_D\bar{\ell} \nu_R H$, 
together with a conserved $U(1)_L$ or $U(1)_{B-L}$ symmetry. However such an approach involves tiny Yukawa couplings $y_D$ a million times smaller that that of the electron. There have been many attempts which yield Dirac neutrinos without relying on such tiny Yukawa couplings
\cite{Arkani-Hamed:2000oup,Murayama:2002je,Thomas:2006gr,Gu:2006dc,Gu:2007gy,Langacker:2011bi,Memenga:2013vc,Chen:2012baa,Farzan:2012sa,Ding:2013eca,Aranda:2013gga,Wang:2016lve,Kanemura:2016ixx,Fujimoto:2016gfu,Valle:2016kyz,Ma:2016mwh,Borah:2017leo,Wang:2017mcy,Bolton:2019bou,Saad:2019bqf,Earl:2019wjw}. Each of these mechanisms has its own experimental implications, but most studies have not considered GW signatures. One particularly attractive idea is to invoke a Dirac seesaw mechanism as an ultraviolet completion of the dimension-five operator $\frac{1}{\Lambda}\bar{\ell} \nu_R H \sigma$, with both tree-level and one-loop realizations, where $\sigma$ is a scalar and $\Lambda$ denotes the scale of heavy intermediate particles.{\footnote{The Dirac seesaw mechanism, whose minimal version requires a $Z_2$ symmetry, has the phenomenological advantage over just the inclusion of Dirac mass terms in that it can facilitate leptogenesis in its minimal 
realization~\cite{Barman:2022yos}. Furthermore, the particles mediating the one-loop realizations of the Dirac seesaw operator can be dark matter candidates \cite{Yao:2018ekp}, analogous to the scotogenic model \cite{Ma:2006km} in the case of a Majorana seesaw. These are in addition to the theoretical motivation of explaining the smallness of the neutrino Dirac Yukawa couplings, which is not accounted for in the usual Dirac mass model (without a $Z_2$ symmetry).}} Although there are three types of 
Dirac seesaw mechanisms, they all have one thing in common in their minimal formulations, namely they rely on a spontaneously broken $Z_2$ symmetry {under which $\sigma$ and $\nu_R$ are odd}~\cite{Yao:2018ekp}. This leads to domain wall formation, and, including a soft $Z_2$ breaking term, domain wall annihilation and GW production, providing a distinctive signature generic to all Dirac seesaw mechanisms. {Intriguingly, the GW spectrum is determined only by the potential of $\sigma$, thus yielding \emph{identical} signals for \emph{all} Dirac seesaw mechanisms provided $\sigma$ does not couple to a scalar mediator.}

 Motivated by the above considerations, and generic assumptions, it seems possible that the GW spectrum can distinguish the Dirac from Majorana seesaw mechanisms, the former yielding a peaked spectrum (from the $Z_2$ domain walls) and the latter a flat spectrum (from the $U(1)_L$ or $U(1)_{B-L}$ cosmic strings). For definiteness, we focus on the type-I seesaw mechanism for the Majorana mass generation, whereas for Dirac mass generation, without loss of generality we focus on the type-I Dirac seesaw mechanism, since the GW signatures are identical for all Dirac seesaw types.\footnote{We note that the Yukawa couplings which give rise to the heaviest Dirac neutrino mass (around $0.1$ eV) will be required to be of similar magnitude to the third family SM Yukawa couplings, and are thus required to be in the approximate range $0.01-1$, up to a factor of a few. The 
seesaw-like mechanism will also be required to be a self-consistent effective field theory valid up to the highest explicit mass scale appearing in the model, which could be as high as the Planck scale.}

{\blue{An illustrative example of this idea is to analyze the spectral shape of the GW signal at nHz frequencies in the two types of neutrino mass models, where several pulsar timing arrays (PTAs) including NANOGrav~\cite{NANOGrav:2023gor, NANOGrav:2023hvm, NANOGrav:2023ctt, NANOGrav:2023hfp, NANOGrav:2023hde, NANOGrav:2023icp, NANOGrav:2023pdq, NANOGrav:2023tcn}, EPTA~\cite{EPTA:2023sfo, EPTA:2023akd, EPTA:2023fyk, EPTA:2023gyr, EPTA:2023xxk, EPTA:2023xiy}, PPTA~\cite{Zic:2023gta, Reardon:2023gzh, Reardon:2023zen}, and CPTA~\cite{Xu:2023wog} have reported convincing evidence of a stochastic gravitational wave background which cannot arise from a population of inspiraling supermassive black hole binaries. We show that the Majorana mass model which yields a cosmic string induced GW signal is unlikely to produce such a signal, whereas the domain wall induced GW signal from the Dirac mass model remarkably matches the PTA result.}}

The organization of the paper is as follows. In section \ref{sec:models} we discuss models of neutrino mass generation in the Majorana and Dirac cases. Sections \ref{sec:CSGW} and \ref{sec:DWGW} focus on the production of gravitational waves, specifically from cosmic strings in the context of Majorana mass generation, and from domain walls in relation to Dirac mass generation. The resulting signals are examined in section \ref{sec:results}, followed by concluding remarks in section \ref{sec:conclusion}. 

\section{Neutrino masses via the seesaw mechanism} 
\label{sec:models}
In this section we explore the generation of Majorana and Dirac neutrino masses via dimension-five operators that realize the respective seesaw mechanism. Both operators produce small neutrino masses without assuming tiny Yukawa couplings. In both cases a family of tree-level models emerges which share a common theme. For the Majorana seeesaw, the $B-L$ symmetry must be broken, which we assume to be spontaneously broken from an ultraviolet $U(1)_{B-L}$ symmetry. For the Dirac seesaw, $U(1)_{B-L}$ must remain unbroken, but naturally preventing tree-level Dirac mass necessitates a $Z_2$ symmetry, which must be spontaneously broken in the process of generating the Dirac mass of the SM neutrinos. A key observation is that the respective breaking of symmetries in the Majorana and Dirac seesaws, regardless of the specific tree-level model, yields strikingly different GW signatures.

\subsection{Majorana seesaw}


As an example of a seesaw model with a spontaneously broken $U(1)_{B-L}$ symmetry, we consider a type-I seesaw in which the SM is extended with three right-handed neutrinos $\bar{N}_i$ and a scalar $\phi$, both singlet under the SM gauge groups. However, the model has an anomaly-free gauged $B-L$ symmetry, under which $\phi$ has two units of charge and $\bar{N_i}$ have a single unit of charge. 
The Lagrangian of the model is given by
\begin{align}
	-\mc L_{M} \supset \mc Y\ \bar{\ell}H\bar{N} + \bar{N} \bar{N}^{T} \phi\,,
\end{align}
which yields the diagram,
\begin{equation*}\label{Majorana_seesaw}
	\centering
	\begin{tikzpicture}[baseline=(a.base)]
	\begin{feynman}[small]
	\vertex (a);
	\vertex [right=of a] (b);
	\vertex [right=of b] (c);
	\vertex [below=of b] (d) {\(\phi\)};
	\vertex [above left=of a] (i1) {\(\bar{\ell}\)};
	\vertex [below left=of a] (i2) {\(H\)};
	\vertex [above right=of c] (f1) {\(\bar{\ell}\)};
	\vertex [below right=of c] (f2) {\(H\)};
	\diagram* {
		(i1) -- (a) --[edge label=\(\bar{N}\quad \bar{N}\)] (c) -- (f1),
		(a) -- [scalar] (i2),
		(b) -- [scalar] (d),
		(c) -- [scalar] (f2),
	};
	\end{feynman}
	\end{tikzpicture}\,.
\end{equation*}
The right-handed neutrinos acquire heavy Majorana masses after the $B-L$ symmetry is spontaneously broken when $\phi$ gets a nonzero vacuum expectation value (VEV). 
Light neutrino masses are generated by integrating out the heavy right-handed neutrinos, and their mass matrix is given by 
\begin{align}
	\mc M_M = \frac{1}{\sqrt{2}} v^2\ \mc Y\ \mc M_{N}^{-1}\ \mc Y^{T}\,,
\end{align}
where $\mc M_N$ is the mass matrix of the right-handed neutrinos.

The most elegant type-I seesaw models invoke a gauged $U(1)_{B-L}$ instead of a bare mass term for the right-handed neutrinos explicitly breaking the $B-L$ symmetry. This is motivated by anomaly cancellation, separating the scale of the right-handed neutrino masses from the grand unification scale, where the breaking chains may contain an intermediate $U(1)_{B-L}$ in grand unified theories such as $SO(10)$, Pati-Salam and left-right symmetric models.

The main characteristic of this scenario is the breaking of the $U(1)_{B-L}$ symmetry, which creates a cosmic string network that eventually decays and produces a stochastic gravitational wave background. As we will discuss in section \ref{sec:CSGW}, such GW signals are nearly flat for a vast range of observable frequencies in gravitational wave interferometers. We note that there could be a secondary contribution to the GW spectrum if the scalar $\phi$ undergoes a first order phase transition (FOPT) when it spontaneously breaks the $U(1)_{B-L}$ symmetry. However, it is well known that the signal from FOPT of a single scalar is typically suppressed compared to the cosmic string signal, particularly when the $U(1)$ symmetry is broken at a sufficiently high scale. Hence, we will not consider the FOPT signal for this model.

\subsection{Dirac seesaw}
\begin{figure}[t]
    \centering
    \subfloat[type-I\label{fig_Dirac1}]{
        \includegraphics[width=0.27\textwidth]{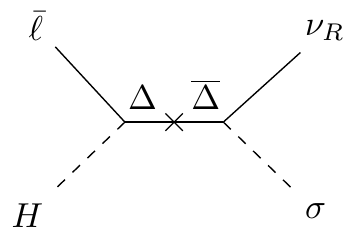}
    }\qquad
    \subfloat[type-II\label{fig_Dirac2}]{
        \includegraphics[width=0.27\textwidth]{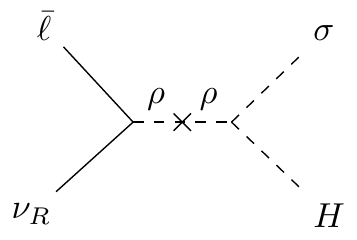} 
    }\qquad
    \subfloat[type-III\label{fig_Dirac3}]{
        \includegraphics[width=0.27\textwidth]{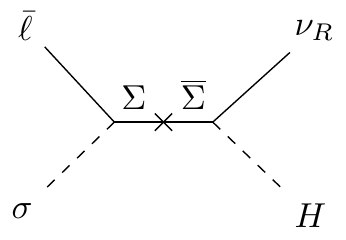} 
    }
\caption{Three tree-level realizations of the dimension-five Dirac seesaw operator $\bar{\ell}\nu_R H \sigma$. A $Z_2$ symmetry is required in all three cases to forbid a tree-level Dirac mass term $\bar{\ell}\nu_R H$.
} 
\label{fig:DiracFeynman}
\end{figure}

Analogous to the Majorana seesaw, there are three tree-level realizations of the Dirac seesaw operator $\bar{\ell}H \nu_R \sigma$ depending on the mediator field. In Fig.~\ref{fig:DiracFeynman} we show the Feynman diagrams corresponding to these cases.

The intermediate Dirac fermion $\Delta$ is a singlet under SM $SU(2)_L$, and the intermediate scalar $\rho$ and fermion $\Sigma$ transform as doublets. In all three cases the scalar $\sigma$ and the right-handed neutrino $\nu_R$ are gauge singlets. 

The Dirac-ness of the neutrino mass can be ensured by imposing a $U(1)_L$ or $U(1)_{B-L}$ symmetry which remains unbroken. It was noted in Ref.~\cite{Yao:2018ekp} that 
the Dirac mass generated via any of the three diagrams is guaranteed to be the leading contribution by imposing a $Z_2$ symmetry under which $\nu_R$ and $\sigma$ are odd. This statement holds even for one-loop realizations of the Dirac seesaw operator \citep{Yao:2018ekp}.
The essential point is that all minimal ultraviolet completions of the operator $\bar{\ell} \nu_R H \sigma$ at the tree and one-loop level requires a $Z_2$ symmetry, which is spontaneously broken when $\sigma$ gets a nonzero VEV. 

{As a concrete example, we will discuss the type-I scenario for the remainder of the paper, but we emphasize that our main argument, i.e. spontaneous breaking of the $Z_2$ symmetry, would be a common feature of all relevant scenarios.}  
The Lagrangian of the type-I model is given by
\begin{align}
	-\mathcal{L}_D \supset \mathcal{Y}_L \bar{\ell} H \Delta_R + \mathcal{Y}_R \overline{\Delta}_L \sigma \nu_R + \mathcal{M}_{\Delta} \overline{\Delta}{\Delta}\,. \label{eq:Lagrangian}
\end{align}
We assume that the Dirac fermion $\Delta$ is heavy. After integrating it out, when the SM Higgs and the new scalar get VEVs $v$ and $u$, respectively, an effective Dirac mass term $\mathcal{M}_D \bar{L} \nu_R$ for the light neutrinos is generated, where 
\begin{align}
	\mathcal{M}_D = \frac{1}{\sqrt{2}}\ v\ u\ \mathcal{Y}_L \mathcal{M}_{\Delta}^{-1} \mathcal{Y}_R \label{DiracMass}
\end{align}
is the Dirac mass matrix suppressed by the large eigenvalues of the mass matrix $\mathcal{M}_{\Delta}$ of the heavy fermions $\Delta$.

The scalar $\sigma$ spontaneously breaks the $Z_2$ symmetry when it acquires a nonzero VEV, necessary for Dirac mass generation. This leads to the creation of domain walls. Long-lived domain walls are dangerous for cosmology if they dominate the energy density of the Universe. However, they can be made to annihilate into gravitational waves by softly breaking the $Z_2$ symmetry, which lifts the degeneracy between the two $Z_2$ vacua. This leads to characteristic GW signals peaked at a single frequency. Since the global lepton number symmetry remains unbroken due to the Dirac nature of the neutrinos, this setup does not lead to the generation of cosmic strings and an associated flat GW spectrum as in the Majorana case. 


\section{GWs from cosmic strings in Majorana seesaw model} \label{sec:CSGW}

To make the discussion self-contained, we briefly recount how GWs are produced from the spontaneous breaking of a gauged $U(1)$ symmetry. The breaking of the $U(1)_{B-L}$ symmetry leads to the creation of a horizon-length string network \cite{Kibble:1976sj}. We specifically focus on Nambu-Goto cosmic strings that lose energy primarily through loop formation and emission of gravitational radiation. The energy density in the string network is diluted by producing closed string loops \cite{Vilenkin:2000jqa}, about $10\%$ of which are large loops and the remaining are highly boosted smaller loops \cite{Vanchurin:2005pa, Olum:2006ix, Martins:2005es, Ringeval:2005kr}. The formation of the loops from long string networks can be described using the velocity-dependent one-scale model \cite{Martins:1995tg, Martins:1996jp}. The loop formation rate is assumed to be equal to the rate of energy loss of the evolving long string network in a cosmological background, and is given by
\begin{align}
    \frac{d n_\alpha}{dt} = \mc F_a \frac{C_{\rm eff}}{\alpha}\frac{1}{t^4}\,, \label{loop}
\end{align}
with the parameter values $\alpha \simeq \mc F_{\alpha} \simeq 0.1$, $C_{\rm eff} \simeq 0.5$ and $5.7$ during matter and radiation domination, respectively, are found from lattice simulations \cite{Blanco-Pillado:2017oxo}. 

While the kinetic energy of the smaller loops are diluted by simple redshifting, the larger loops oscillate and emit energy in the form of gravitational waves at a constant rate,
\begin{align}
    \frac{dE}{dt} = -\Gamma G \mu^2\,, \label{energy}
\end{align}
where $\Gamma \simeq 50$ is a dimensionless constant \cite{Vilenkin:1981bx}, $G$ is the Newton's constant and $\mu$ is the tension in the strings. Typically $\mu \sim \mc O(\Lambda)$, where $\Lambda$ is the scale of the $U(1)$ symmetry breaking. As a consequence of emitting gravitational radiation, the initial length of a large loop created by
the network at time $t_i$, given by $l_i = \alpha t_i$, decreases as 
\begin{align}
    \ell (t) = \alpha t_{i} - \Gamma G \mu (t-t_i)\,, \label{length}
\end{align}
until the loop completely disappears.
The total energy loss from a loop can be decomposed into normal modes with frequency $\tilde{f}_k = 2k/\ell$ at a time $\tilde{t}$, where $k=1,2,3, \ldots$ is the mode number. Accounting for redshift evolution, the frequency today becomes $f = [a(\tilde{t})/a(t_0)]\ \tilde{f}_k$, where $t_0$ is the current time. The relative emission rate per mode is given by
\begin{align}
    \Gamma^{(k)} = \frac{\Gamma k^{-4/3}}{\sum_{j=1}^{\infty} j^{-4/3} } \simeq \frac{\Gamma k^{-4/3}}{3.60}\,. 
\end{align}
Combining Eqs.~\eqref{loop}, \eqref{energy} and \eqref{length}, and integrating over the emission time, the gravitational wave amplitude of the $k$-th mode is given by
\begin{align}
    \Omega_{\rm GW}^{(k)} (f) = \frac{1}{\rho_c}\frac{2k}{f} \frac{\mc F_\alpha \Gamma^{(k)}G\mu^2}{\alpha (\alpha + \Gamma G\mu)} \int_{t_F}^{t_0} d\tilde{t} \frac{C_{\rm eff} (t_i^{(k)})}{{t_i^{(k)}}^4} \left[ \frac{a(\tilde{t})}{a(t_0)} \right]^5 \left[\frac{a(t_i^{(k)})}{a(\tilde{t})}\right]^3\ \Theta(t_i^{(k)} - t_F)\,, \label{GWCS}
\end{align}
where $\rho_c = 3H_0^2/{(8\pi G)}$ is the critical energy density,  $t_i^{(k)}$ is the formation time of loops contributing to the $k$-th mode and is given by
\begin{align}
    t_i^{(k)}(\tilde{t}, f) = \frac{1}{\alpha + \Gamma G \mu}\left[ \frac{2k}{f} \frac{a(\tilde{t})}{a(t_0)} + \Gamma G \mu \tilde{t} \right]. 
\end{align}
Summing over all modes, we get the total amplitude of the gravitational waves
\begin{align}
    \Omega_{\rm GW} (f) = \sum_k \Omega_{\rm GW}^{(k)} (f)\,,
\end{align}
where the sum can be easily evaluated using 
\begin{align}
    \Omega_{\rm GW}^{(k)} (f) = \frac{\Gamma^{(k)}}{\Gamma^{(1)}} \Omega_{\rm GW}^{(1)}(f/k) = k^{-4/3}\ \Omega_{\rm GW}^{(1)} (f/k)\,.
\end{align}

\section{GWs from domain walls in Dirac seesaw model} \label{sec:DWGW}
We now focus on the GW spectrum generated in the type-I realization of the Dirac seesaw operator. We will argue that the GW spectrum is identical for any tree-level or one-loop realization of the Dirac seesaw operator in which the scalar $\sigma$ does not mix with the SM Higgs or other scalars.

We assume a simple potential for  $\sigma$:
\begin{align} \label{V0dirac}
	V(\sigma) = \frac{\lambda}{4}(\sigma^2 - u^2)^2\,.
\end{align}
This potential has two degenerate minima at $\sigma = \pm u$ and is symmetric under a $Z_2$ transformation $\sigma \rightarrow -\sigma$. Domain walls are formed around the boundaries of these two minima.
The symmetry is spontaneously broken when the scalar chooses one of the two vacua. 
This choice depends on random fluctuations of the field and is made independently at spatially distant regions in space, creating the so-called `domains'. 
Domain walls are formed around the boundaries of these  domains.
We assume that the domain walls have a static planar configuration perpendicular to the $z$ direction. Introducing a kinetic term $\frac{1}{2} (\partial_\mu \sigma)^2$, the field equation for $\sigma(z)$ is given by
\begin{align}
    \frac{d^2\sigma}{dz^2} - \frac{dV}{d\sigma} = 0\,, \label{EOMphase}
\end{align}
which yields the solution,
\begin{align}
    \sigma(z) = u\ {\rm tanh}{\left( 
    \sqrt{\frac{\lambda}{2}} u z \right)}\,, \label{sigmasol}
\end{align} 
for the boundary condition $\sigma(z\rightarrow \pm \infty) \rightarrow \pm u$. The surface energy density (also called tension) of the wall can be derived from integrating the $00$ component of the stress-energy tensor $\mc T_{\mu \nu} = (d\sigma/dz)^2\ \rm{diag}(+1, -1,-1, 0)$, and is given by
\begin{align}
	\mc E = \frac{2}{3}\sqrt{2 \lambda}\ u^3\,.
\end{align}

Domain walls can be very long-lived and may dominate the energy density of the Universe, alter its equation of state and lead to rapid expansion inconsistent with standard cosmology. Even if their energy density is subdominant today, domain walls may produce excessive density perturbations observable in the CMB epoch if their surface energy density is above $\mc O({\rm MeV}^3)$~\cite{Zeldovich:1974uw}. 

An interesting solution to the domain wall problem is to softly break the discrete symmetry that lifts the degeneracy between the vacua. We introduce an explicit breaking term in the potential,
\begin{align} \label{dVdirac}
	\Delta V(\sigma) = \epsilon u \sigma \left(\frac{\sigma^2}{3} - u^2\right)\,,
\end{align}
where $\epsilon$ is a dimensionless constant. The overall potential $V(\sigma) + \Delta V(\sigma)$ still has two minima at $\sigma = \pm u$, but with a difference in the potential at these points:
\begin{align}
	V_{\rm bias} = V(-u) - V(+u) = \frac{4}{3}\epsilon u^4\,.
\end{align}
The probability $p_-$ of a domain ending up in the $-u$ vacuum (`false' vacuum) is smaller compared to $p_+$ of it being in the $+u$ vacuum (`true' vacuum), their ratio being related to the potential difference
\begin{align}
	\frac{p_-}{p_+} \simeq \exp{\left(-\frac{V_{\rm bias}}{V_0}\right)}\,,
\end{align} 
where 
\begin{align}
    V_0 = \frac{u^4}{12\lambda^3}(3\lambda - \epsilon)(\lambda+\epsilon)^3 \label{V0def}
\end{align}
is the potential difference between the maximum and the $+u$ minimum. Treating the system as a three-dimensional lattice, percolation theory predicts that an infinite cluster of the false vacuum appears in space if the corresponding probability is above the threshold $p_c = 0.311$ \cite{Stauffer:1978kr}. This yields an upper bound on the bias potential for the generation of domain walls, $V_{\rm bias} < V_0 \log{\frac{1-p_c}{p_c}} = 0.795 V_0$,
which can be written as
\begin{align}
    V_{\rm bias} < 0.38 \lambda u^4. \label{constraint:C} 
\end{align}
As long as the bias potential is below this limit, domain walls are created and their dynamics is mostly controlled by the surface energy density. The energy density of the wall in this regime is given by a scaling solution \cite{Press:1989yh},
\begin{align}
	\rho_{\rm wall}(t) = \mc A \frac{\mc E}{t}\,,
\end{align}
where $\mc A \simeq 0.8 \pm 0.1$ is the so-called area parameter \cite{Hiramatsu:2013qaa}.
A volume pressure $p_v \sim V_{\rm bias}$ tends to shrink the false vacuum region. Domain walls collapse when the volume pressure overcomes the pressure from surface energy density, which happens at
\begin{align}
	t_{\rm ann} = \mc C_{\rm ann} \mc A \frac{\mc E}{V_{\rm bias}}\,,
\end{align}
where $\mc C_{\rm ann} = 5$ for $Z_2$ breaking \cite{Kawasaki:2014sqa}.

Assuming that the domain walls annihilate during the radiation dominated era and annihilation happens instantaneously at $t=t_{\rm ann}$, the peak amplitude of the generated gravitational waves at present time $t_0$ can be expressed as \cite{Saikawa:2017hiv}
\begin{align}\label{omegapeak}
	\Omega^{\rm peak}_{\rm GW}h^2 (t_0) &\simeq 1.49 \times 10^{-10} \times \left(\frac{\tilde{\epsilon}_{\rm GW}}{0.7}\right) \left(\frac{\mc A}{0.8}\right)^4 \left(\frac{10.75}{g_\star}\right)^{1/3} \left(\frac{\mc E^{1/3}}{10^7\ \rm GeV}\right)^{12} \left(\frac{10^7\ \rm GeV^4}{V_{\rm bias}}\right)^{2}\,,
\end{align}
and the peak frequency is given by
\begin{align} \label{fpeak}
	f_{\rm peak} \simeq 5.93 \times 10^{-6}\ {\rm Hz} \times  \left(\frac{5}{\mc C_{\rm ann} }\right)^{1/2}\ \left(\frac{0.8}{\mc A}\right)^{1/2} \left(\frac{10^7\ \rm GeV}{\mc E^{1/3}}\right)^{3/2} \left(\frac{V_{\rm bias}}{10^7\ \rm GeV^4}\right)^{1/2}\,,
\end{align}
where the parameter $\tilde{\epsilon}_{\rm GW}$ is estimated to be $\tilde{\epsilon}_{\rm GW} \simeq 0.7 \pm 0.4 $ \cite{Hiramatsu:2013qaa}. 
Here, $g_{*s}$ is the relativistic degrees of freedom for the entropy density at temperature $T$. 
Numerical simulations suggest that the gravitational wave amplitude rises as $\Omega_{\rm GW} \propto f^3$ for $f < f_{\rm peak}$ and falls off as $\Omega_{\rm GW} \propto f^{-1}$ for $f > f_{\rm peak}$. {\blue{Joined by a smoothing function~\cite{Caprini:2019egz},
\begin{align}
    S(x) = \frac{(a+b)^c}{\left(b x^{-a/c}+a x^{b/c}\right)^{c}}\,,
\end{align}
the gravitational wave spectrum can be written as
\begin{align} \label{omegadw}
    \Omega_{\rm GW}^{\rm DW}h^2 (f) \simeq \Omega_{\rm GW}^{\rm peak}h^2\ S\left(\frac{f}{f_{\rm peak}}\right)\,.
\end{align}
Here the low-frequency slope of the signal is required by causality to be $a=3$~\cite{Caprini:2019egz}, and numerical simulations suggest that the high-frequency slope $b$ and the width near the spectral peak $c$ are close to unity~\cite{Hiramatsu:2010yz, Hiramatsu:2013qaa}. Uncertainties in the numerical simulations allow  $0.5 \leq b \leq 1$ and $0.3 \leq c \leq 3$~\cite{Ferreira:2022zzo}.
}}

{It is evident from Eqs.~(\ref{omegapeak})-(\ref{omegadw}) that the GW spectrum depends on $\mc E$ and $V_{\rm bias}$, both of which are determined by the $Z_2$-symmetric and explicit-breaking part of the potential of the scalar $\sigma$. The specific realization of the Dirac seesaw operator does not impact the GW signal {as long as $\sigma$ does not have a mixing term with another scalar in its potential}, although the Dirac masses of the neutrinos do depend on the underlying model \cite{Yao:2018ekp}.}

It should be noted that another possibility for obtaining GWs in this model is through a first order phase transition induced by the scalar $\sigma$. The bias potential in Eq.~\eqref{dVdirac} contains a cubic term which creates a barrier between the true and false vacua at zero temperature. However, we have explicitly checked for $10^{-6} < \lambda < 1$, $10^3<u<10^{15}$ GeV and $10^{-9}<\epsilon<1$ that a FOPT either does not occur or is extremely weak. This is because the linear term in the potential, also controlled by the parameter $\epsilon$, tends to remove the barrier between the two vacua. It is conceivable that in a variation of the model, a FOPT would occur, which would produce a peaked
GW spectrum, more akin to that from domain walls than cosmic strings.

\section{Results} \label{sec:results}
In this section we discuss the gravitational wave signatures from cosmic strings and domain walls in the context of Majorana and Dirac seesaw models, respectively. Existing and planned interferometers probe frequencies from $10^{-9}$ to $10^{4}$~Hz range. In the nanoHz range ($10^{-9} - 10^{-7}$~Hz), currently operating pulsar timing arrays (PTA) EPTA \cite{vanHaasteren:2011ni} and NANOGrav \cite{NANOGrav:2020bcs} have set upper bounds on the stochastic GW background, and the upcoming SKA \cite{Janssen:2014dka} and IPTA \cite{Verbiest:2016vem} interferometers will have much greater sensitivity. $\mu$Ares \cite{Sesana:2019vho} will be sensitive to the $\mu$-Hz to Hz band. The mHz to Hz band will be further probed by future laser interferometers LISA \cite{LISA:2017pwj}, BBO \cite{Harry:2006fi} and DECIGO \cite{Kudoh:2005as, Kawamura:2020pcg}, as well as by atomic interferometers AION \cite{Badurina:2019hst} and AEDGE~\cite{AEDGE:2019nxb}. Around the 100 Hz, Advanced LIGO+VIRGO \cite{LIGOScientific:2022sts}  have set an upper limit \cite{KAGRA:2021kbb, Jiang:2022uxp} and their future upgrades will improve on the sensitivity by at least an order \cite{KAGRA:2021kbb}. Einstein Telescope (ET) \cite{Hild:2008ng} and Cosmic Explorer (CE) \cite{LIGOScientific:2016wof} are planned to operate in the same band with three orders of magnitude greater sensitivity. 

\begin{figure}[t]
    \centering
    \includegraphics[width=0.95\textwidth]{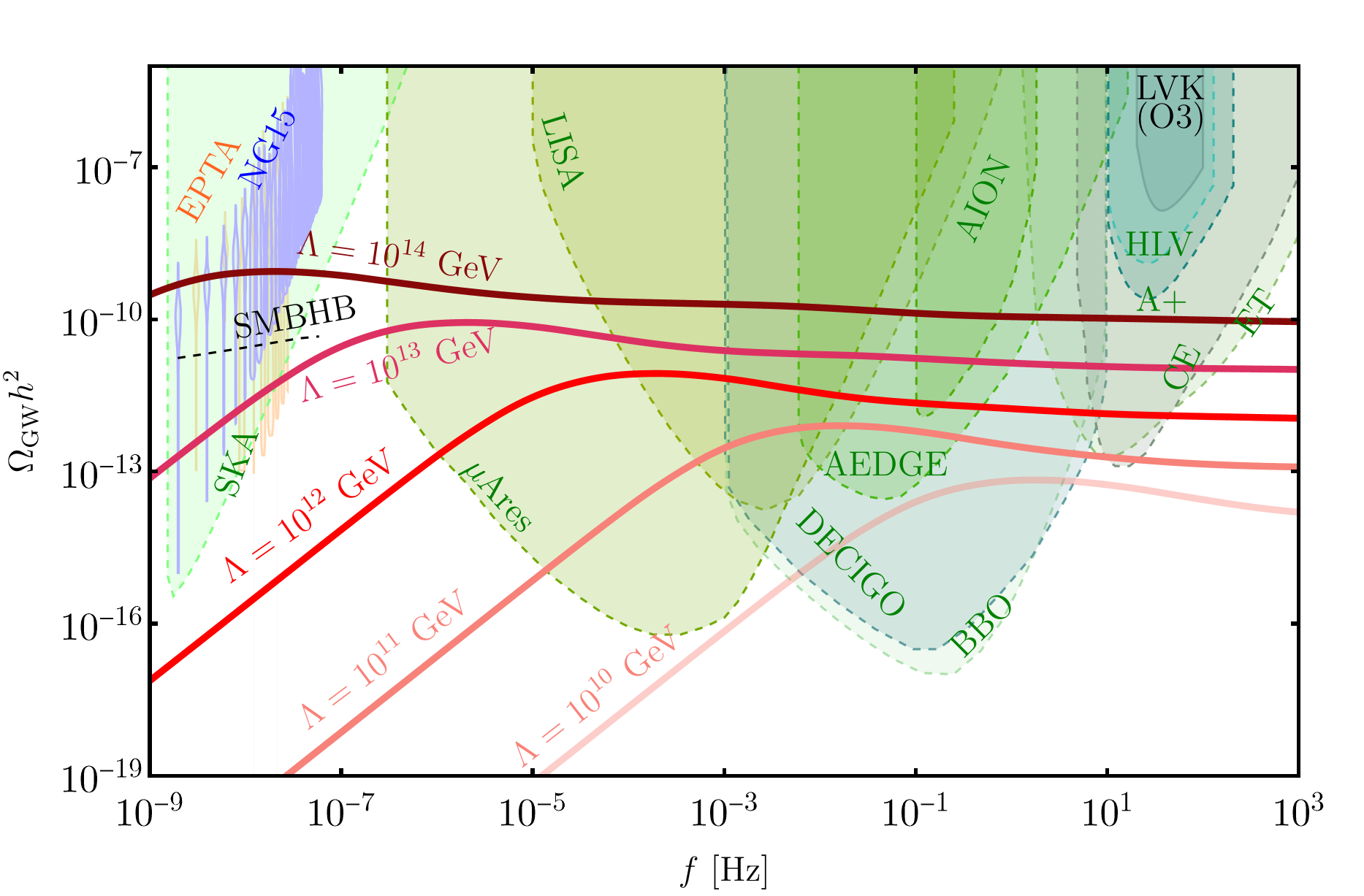}
    \caption{Gravitational wave spectrum induced by cosmic strings generated via the spontaneous breaking of the gauged $U(1)_{B-L}$ symmetry responsible for Majorana mass of the neutrinos. $\Lambda$ denotes the scale of symmetry breaking.}
    \label{fig:GWCS}
\end{figure}

For the Nambu-Goto cosmic string network, the only free parameter in Eq.~\eqref{GWCS} is $\mu$ of $\mc O(\Lambda^2)$, where $\Lambda$ is the scale of the $U(1)_{B-L}$ symmetry breaking that generates the Majorana masses of the right-handed neutrinos. In Fig.~\ref{fig:GWCS} we show the GW spectrum for $\Lambda = 10^{14}, 10^{13}$, $10^{12}$, $10^{11}$ and $10^{10}$ GeV, corresponding to a high scale of the right-handed neutrino masses. For comparison, we also show the sensitivity and upper bounds of various interferometers spanning a large range of frequencies from nano-Hz to kilo-Hz. The characteristic shape of the cosmic string induced GW signal is a rising spectrum at low frequencies which plateaus at higher frequencies. The height of this plateau is proportional to the symmetry breaking scale. The signals for $\Lambda \gtrsim 10^{14}$~GeV are ruled out by EPTA, whereas signals for smaller scales are within the sensitivity of several interferometers.  

For the Dirac seesaw, the parameter space is subject to various physical constraints that impact the formation and stability of the domain walls \cite{Saikawa:2017hiv, Gelmini:2020bqg}.
If the bias potential is sufficiently small, domain walls collapse too late and may dominate the energy density of the Universe. The time at which domain walls become dominant is given by 
\begin{align}
	t_{\rm dom} = \frac{3}{4} \frac{M_{\rm Pl}^2}{\mc A \mc E}\,.
\end{align}
Requiring $t_{\rm ann} < t_{\rm dom}$ yields a lower bound on the bias potential, $V_{\rm bias} > {4 }C_{\rm ann} \mc A^2 \mc E^2/(3 M_{\rm Pl}^2)$, which can be written as
\begin{align}
	V_{\rm bias}^{1/4} > 8.95\times 10^{-10}\ {\rm GeV}\ \lambda^{1/4}\ \left(\frac{\mc C_{\rm ann}}{5}\right)^{1/4} \left(\frac{\mc A}{0.8}\right)^{1/2} \left(\frac{u}{\rm GeV}\right)^{3/2}\,. \label{constraint:A}
\end{align} 
Even if the domain walls decay before they overclose the Universe, their decay products may destroy the light element abundances created by Big Bang nucleosynthesis (BBN). Assuming that a significant fraction of the energy density of the domain walls is converted into energetic particles, constraints on energy injection at the epoch of BBN require $t_{\rm ann} \lesssim t_{\rm BBN}\simeq 0.01\ {\rm sec}$ \cite{Kawasaki:2004yh, Kawasaki:2004qu}, which can be written as
\begin{align}
	V_{\rm bias}^{1/4} > 3.97\times 10^{-6}\ {\rm GeV}\ \lambda^{1/8}\ \left(\frac{\mc C_{\rm ann}}{5}\right)^{1/4} \left(\frac{\mc A}{0.8}\right)^{1/4} \left(\frac{u}{\rm GeV}\right)^{3/4}\,. \label{constraint:B}
\end{align}

\begin{figure}
    \centering
    \subfloat[ \label{fig:lambda1}]{        \includegraphics[width=0.447\textwidth]{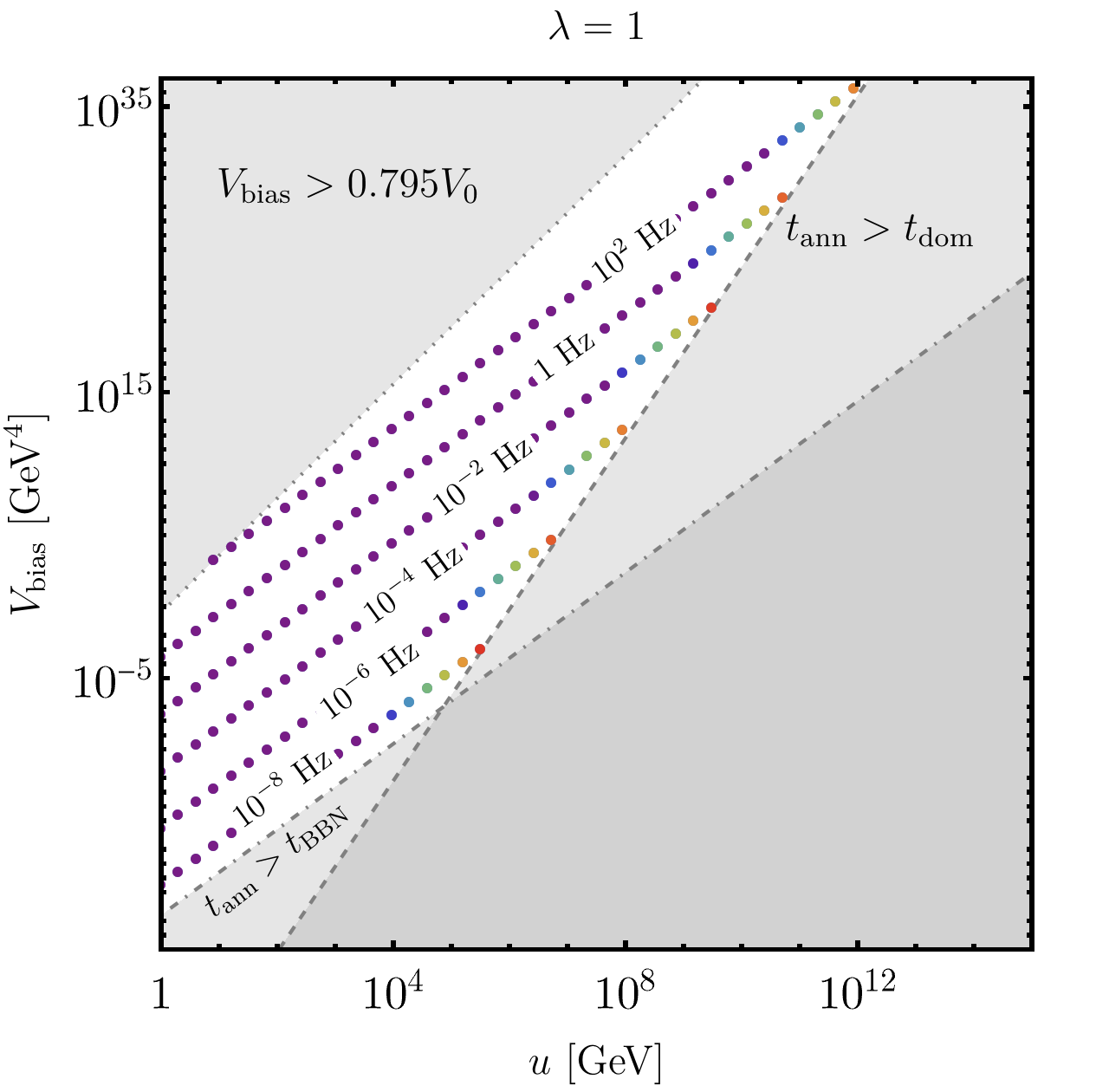}
    }
    \hspace{-0.6cm}
    \subfloat[ \label{fig:lambdam3}]{      \includegraphics[width=0.55\textwidth]{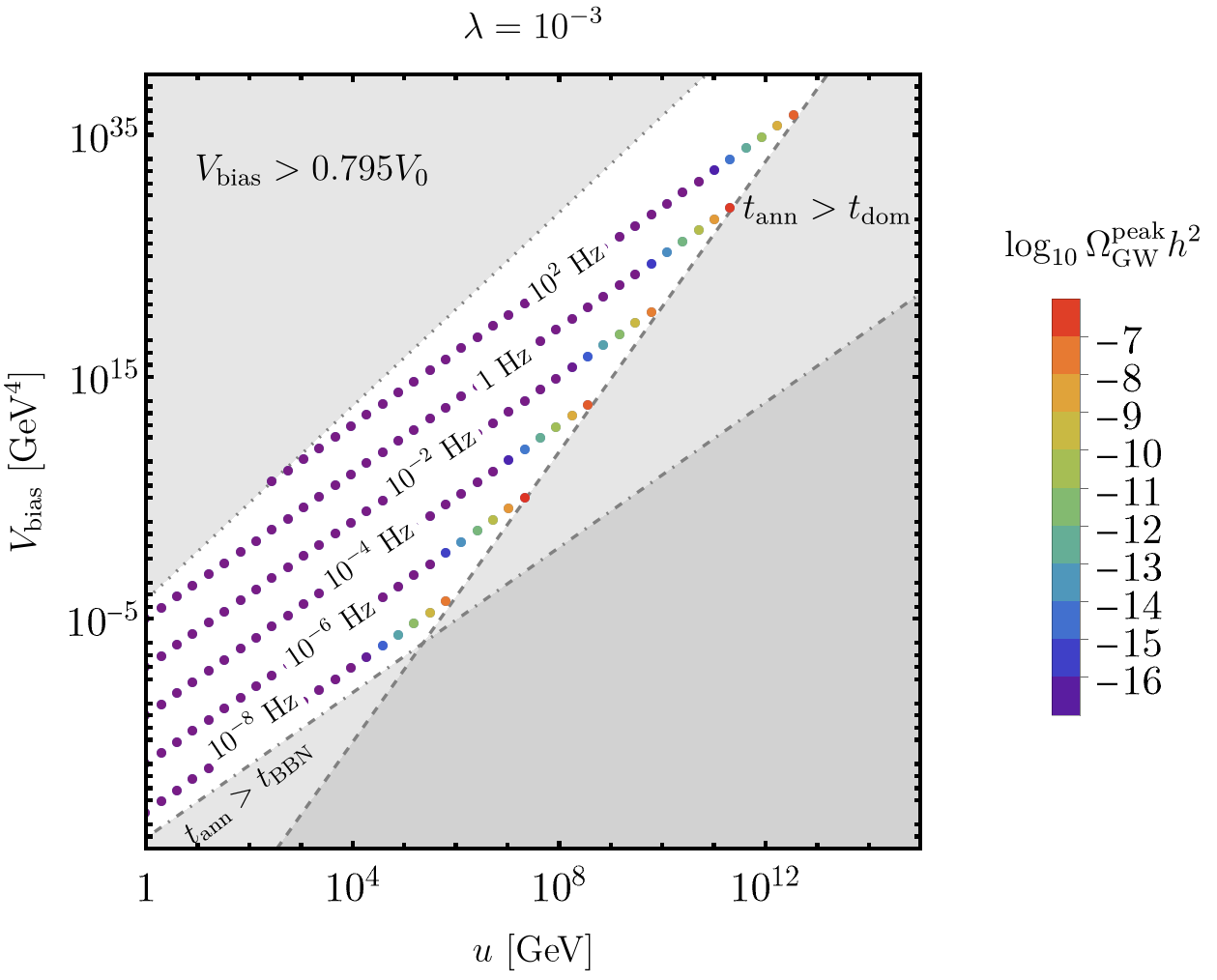} 
    }
    \caption{Parameter space of the Dirac seesaw model for creation and subsequent annihilation of the domain wall network. Gray shaded regions show the parameter space ruled out by physical constraints. Dotted lines in the allowed parameter space represent contours of peak frequency, and the colors of the dots indicate the amplitude of the gravitational wave signal at that frequency.}
    \label{fig:dwparam}
\end{figure}

Equations~\eqref{constraint:A} and \eqref{constraint:B}, together with Eq.~\eqref{constraint:C}, constrain the parameter space for annihilation of domain walls and subsequent gravitational wave production. In terms of the scalar VEV $u$ and bias potential $V_{\rm bias}$, and choosing $\mc A=0.8,\ \mc C_{\rm ann} = 5$, these constraints can be expressed as
\begin{align}
	t_{\rm ann} < t_{\rm dom}: \quad \frac{V_{\rm bias}}{\rm GeV^4} &> 6.42\times 10^{-37}\lambda \left(\frac{u}{\rm GeV}\right)^6\,, \label{consA}\\[1em]
	t_{\rm ann} < t_{\rm BBN}: \quad \frac{V_{\rm bias}}{\rm GeV^4} &> 2.49\times 10^{-22} \sqrt{\lambda} \left(\frac{u}{\rm GeV}\right)^3\,, \label{consB} \\[1em]
	V_{\rm bias} < 0.795V_0: \quad \frac{V_{\rm bias}}{\rm GeV^4} &< 0.38\lambda \left(\frac{u}{\rm GeV}\right)^4\,. \label{consC}
\end{align}
Finally, from Eq.~\eqref{DiracMass}, if we assume that the mediator fermion mass is below the Planck scale $\mc O(10^{19})$ GeV, and the heaviest light neutrino mass is around $\mc O(0.1)$ eV, we find
\begin{align}
    u \lesssim \frac{\mc O(10^{7})\ \text{GeV}}{y^2}\,.
\end{align}
Here we have assumed a single mediator responsible for the $\mc O(0.1)$ eV neutrino mass, and a single coupling $y=y_L=y_R$ associated with it.
For Yukawa couplings $y \gtrsim \mc O(10^{-2})$, this implies an upper bound on the scale of $Z_2$ symmetry breaking, $u \lesssim \mc O(10^{11})$ GeV. 

The constraints of Eqs.~\eqref{consA}-\eqref{consC} are depicted by the gray shaded regions in Fig.~\ref{fig:dwparam} for (a) $\lambda = 1$ and (b) $\lambda = 10^{-3}$. For smaller $\lambda$, the upper left region expands while the other two regions shrink, as expected from Eqs.~\eqref{consA}-\eqref{consC}.
The peak frequencies of the gravitational waves $f_{\rm peak} = 10^{-8}, 10^{-6}, \ldots 10^2$ Hz, are marked by dots. Their colors represent the amplitude of the GW signal at the corresponding peak frequency. We find that amplitudes above $\Omega_{\rm GW}^{\rm peak} h^2 \sim 10^{-6}$ are ruled out by Eq.~\eqref{consA}, while peak frequencies below $f_{\rm peak} \sim 10^{-8}$~Hz are ruled out by Eq.~\eqref{consB}. 
Interestingly, the allowed region can still generate GW signals within the sensitivity of various interferometers. 

\begin{figure}
    \centering
    \includegraphics[width=0.95\textwidth]{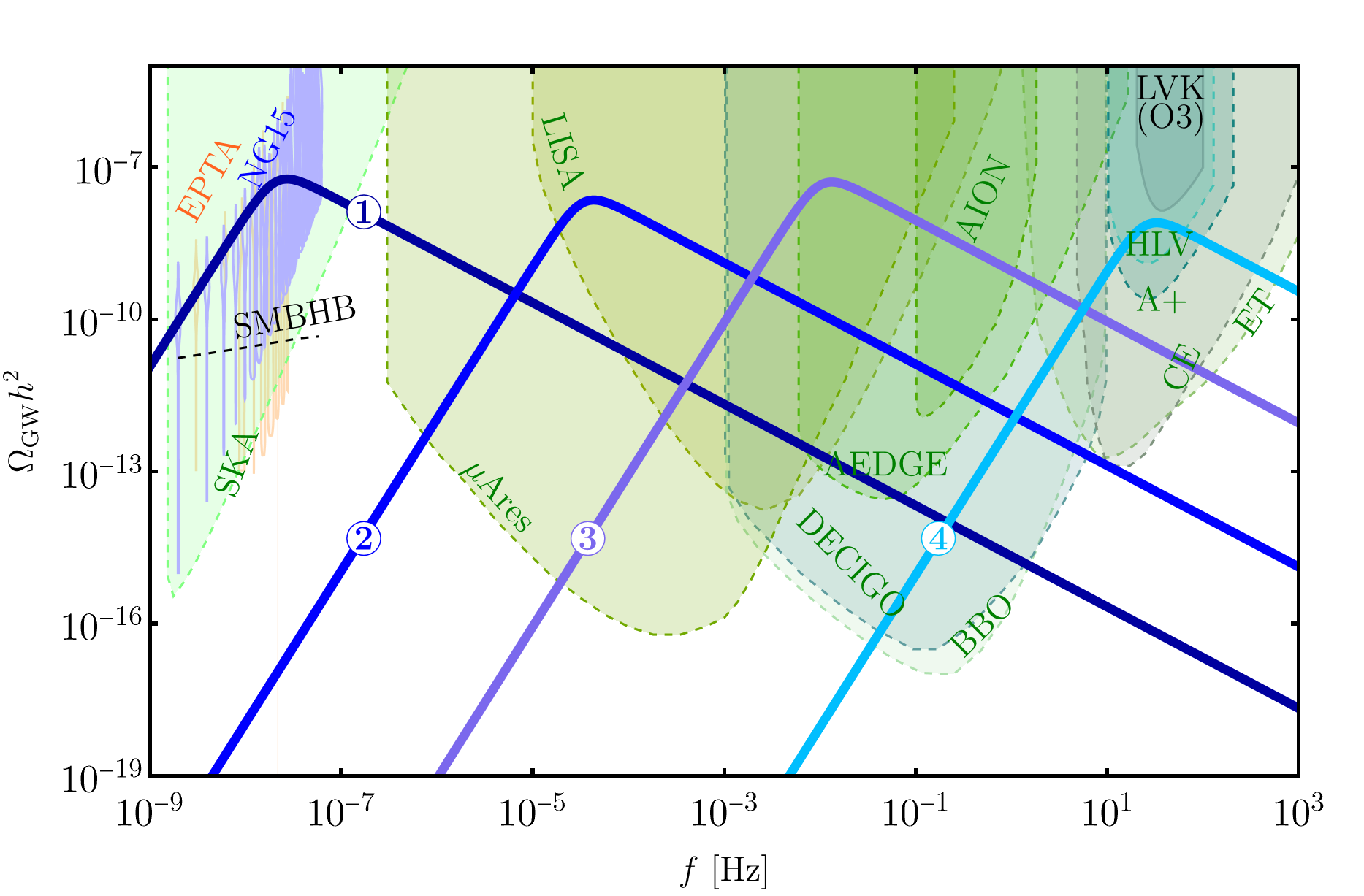}
    \caption{Gravitational wave spectrum from annihilation of domain walls created by soft-breaking of the $Z_2$ symmetry in a Dirac neutrino mass model. Benchmark points \Circled{1}-\Circled{4} are listed in Table~\ref{table:BP}. }
    \label{fig:GWDW}
\end{figure}

\begin{table}
\centering
{\renewcommand{\arraystretch}{1.2}
\begin{tabular}{c@{\hskip 0.3in}c@{\hskip 0.5in}c@{\hskip 0.3in}c}
    \toprule

  Benchmark Point  & ${u}\ [{\rm GeV}]$ & $V_{\rm bias}$ [${\rm GeV}^4$] & $y_{\rm max}(M_{\Delta} < M_{\rm Pl})$ \\
 \midrule
 \Circled{1} & $4.47\times 10^5$ & $1.78 \times 10^{-2}$ & $3.96$\\
 \Circled{2} & $5.2\times 10^{7}$ & $7.14\times 10^{10}$ &   $0.37$\\
\Circled{3} & $2.7 \times 10^{9}$ & $9.3\times 10^{20}$  & $0.051$\\
\Circled{4} & $3.63\times 10^{11}$ & $1.38\times 10^{34}$ &  $0.004$\\
\bottomrule
\end{tabular}
}
\caption{Benchmark points for gravitational wave signals from domain walls with $\lambda = 1$.}
\label{table:BP}
\end{table}

The four benchmark points listed in Table~\ref{table:BP} are chosen from the allowed parameter space. The last column of the table gives the upper bound on the Yukawa coupling, assuming that the mediator mass lies below the Planck scale. We note that the values of such Yukawa couplings cover the range of the third family charged fermion Yukawa couplings in the SM, and only exceed this range by less than an order of magnitude.

\begin{figure}
    \centering
    \includegraphics[width=0.95\textwidth]{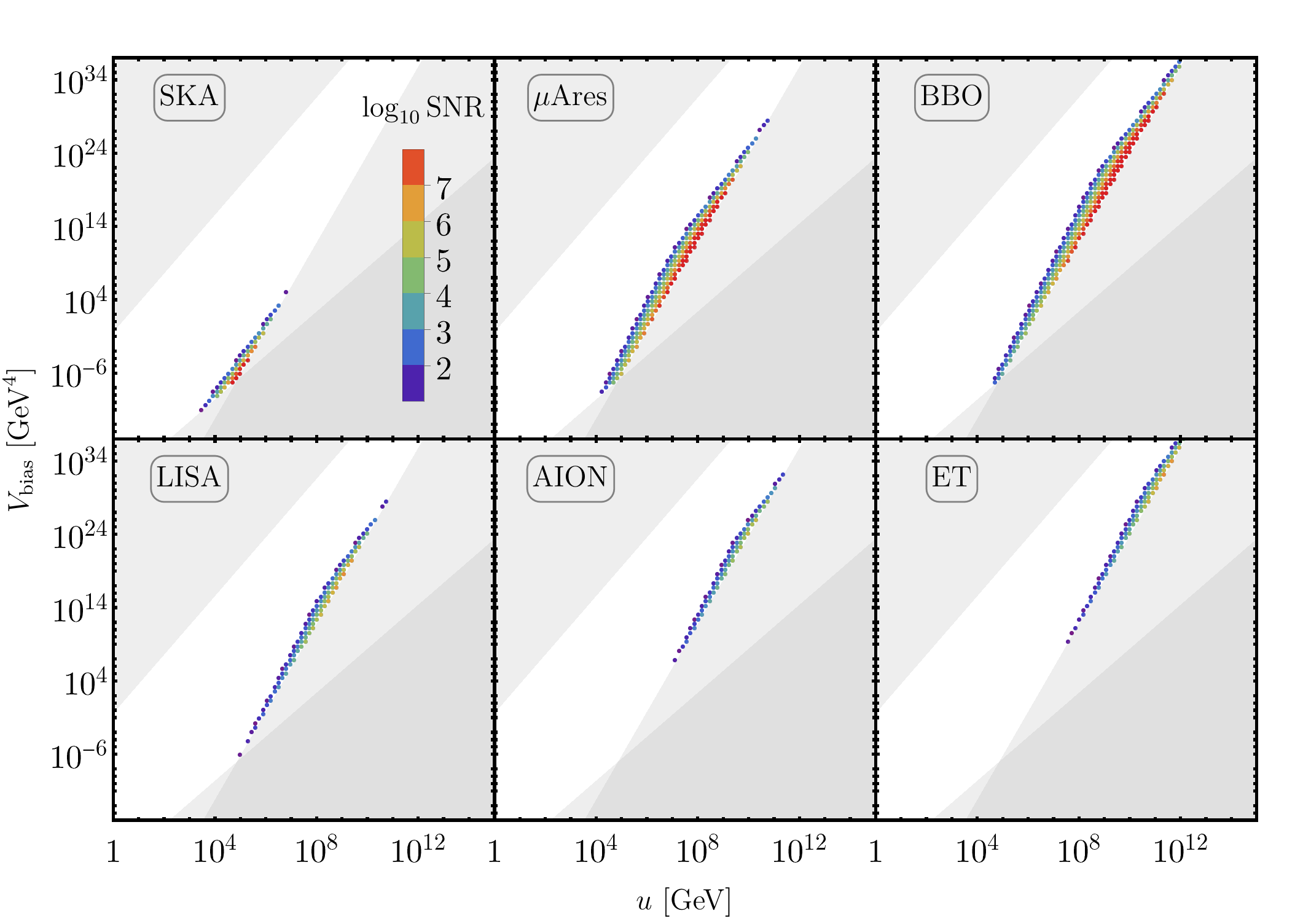}
    \caption{\blue{Signal-to-noise ratio of the gravitational wave signal in the Dirac mass model (setting $\lambda = 1$) at some interferometers. Gray regions are in conflict with the constraints shown Fig.~\ref{fig:lambda1}. Colored regions represent $\text{SNR}\geq 10$.}}
    \label{fig:SNRDW}
\end{figure}

The gravitational wave spectra for these benchmark points are shown in Fig.~\ref{fig:GWDW}. Benchmark point \Circled{1} can be probed by SKA, while \Circled{2} and \Circled{3} 
can be probed by $\mu$Ares, LISA, AEDGE, DECIGO, BBO,
AION, and \Circled{4} by Advanced LIGO+VIRGO, ET and CE, among others. {\blue{We set $b=c=1$ for the spectral shape, but slightly different values still yield a peaked spectrum for $a=3$. In Fig.~\ref{fig:SNRDW}, we show the signal-to-noise 
ratio~\cite{Thrane:2013oya}
\begin{align}
    \text{SNR} \equiv \sqrt{\tau \int_{f_{\rm min}}^{f_{\rm max}} \text{d}f \left[\frac{\Omega_{\rm GW}(f)h^2}{\Omega_{\rm exp}(f)h^2 }\right]^2}\,,
\end{align}
of the generated GW spectrum in the parameter space of the model for various interferometers operating from nHz to kHz. Here $\Omega_{\rm exp}(f)h^2$ is the effective strain noise power spectral density \cite{Moore:2014lga}, $\tau = 4$ years is the typical observation time, and $f_{\rm min}$ and $f_{\rm max}$ define the range of frequencies in which an interferometer is sensitive. We only show the parameter space for $\text{SNR}\geq 10$, which is the threshold for detection in individual interferometers. The results show that detectable signals arise if the domain walls annihilate when they are close to dominating the energy density of the Universe.
}}

The main difference between the signals for the Majorana seesaw model and the Dirac seesaw model is their spectral shape. While cosmic string signals for the former are mostly flat for observable frequencies, domain wall signals for the latter are peaked. We expect that cosmic string signals should be detected at multiple interferometers at different frequency bands, whereas domain wall signals are likely to be detected in only a narrow frequency range. Such a detection will provide valuable information about the nature of neutrino mass genesis and will complement results from neutrinoless double beta experiments.

\subsection{Implications of NANOGrav data}
\begin{figure}
    \centering
    \includegraphics[width=0.355\textwidth]{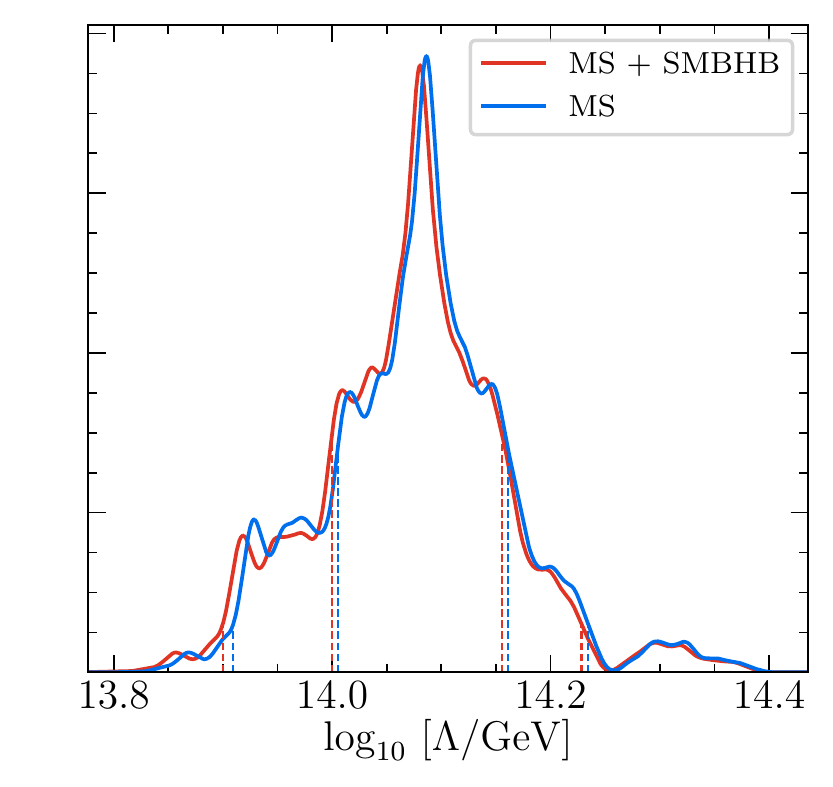}
    \includegraphics[width=0.635\textwidth]{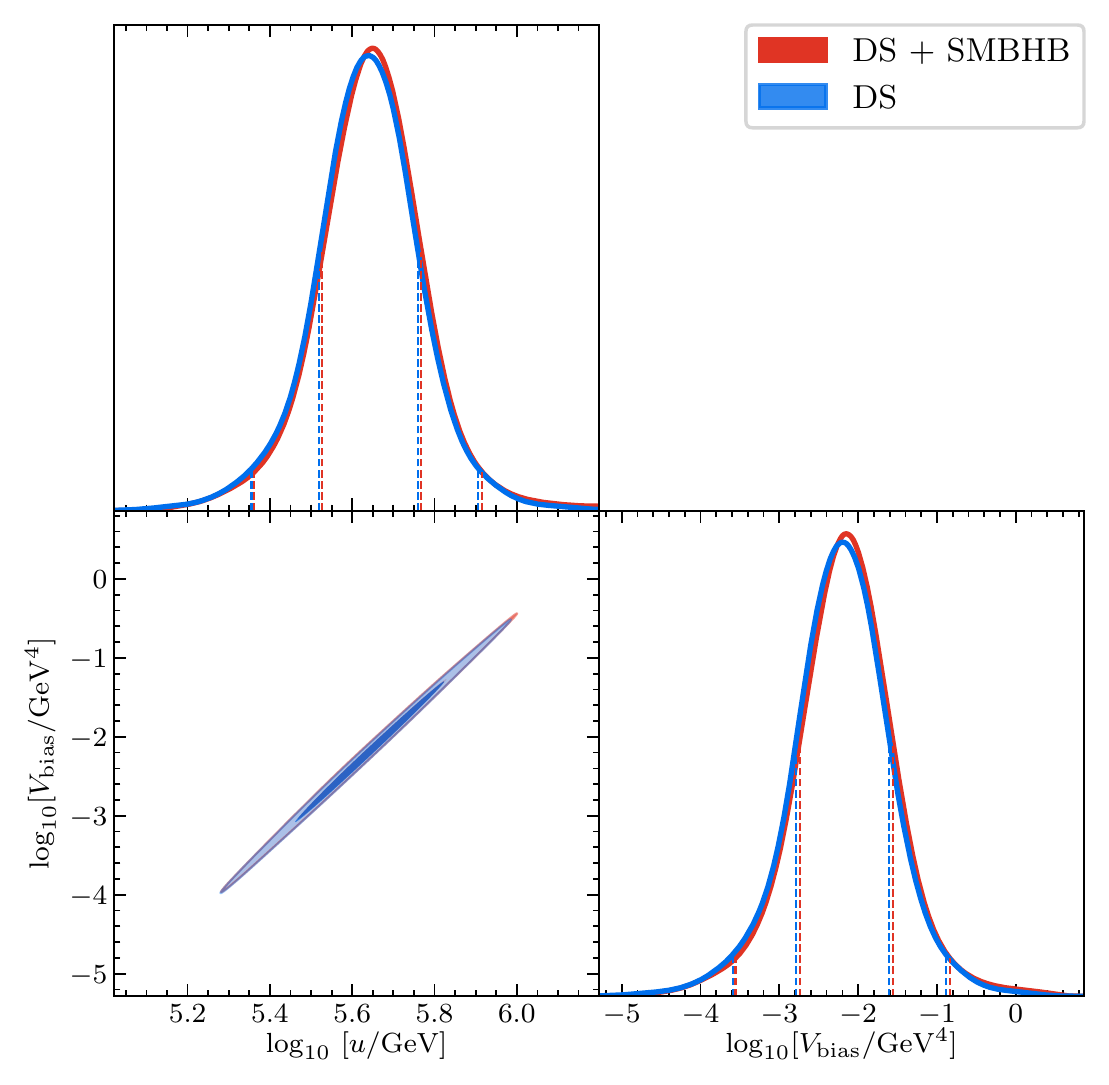}
    
    \caption{Posterior distributions for the Majorana seesaw (MS) model parameter $\Lambda$ (\emph{left}) and the $1\sigma$ and $2\sigma$ credible regions for the Dirac seesaw (DS) model parameters $u$ and $V_{\rm bias}$ (\emph{right}). The blue curves correspond to the GW signal from the seesaw mechanism alone, and the red curves correspond to the combined signal from the seesaw mechanism and supermassive black hole binaries (SMBHBs) {with
     $A_{\rm BHB} = 10^{-15.6}$ and $\gamma_{\rm BHB} = 4.7$.}}
    \label{fig:posteriorDWNG15}
\end{figure}

Finally, we comment on the implications of the recent PTA signals on our Majorana and Dirac seesaw models. For concreteness, we compare the predictions of the models to the NANOGrav 15-year dataset~\cite{NANOGrav:2023gor, NANOGrav:2023hvm, NANOGrav:2023ctt, NANOGrav:2023hfp, NANOGrav:2023pdq, NANOGrav:2023hde, NANOGrav:2023tcn, NANOGrav:2023icp} using the \texttt{PTArcade} software~\cite{Mitridate:2023oar,Lamb:2023jls}. In Fig.~\ref{fig:posteriorDWNG15} we show the posterior distributions and allowed regions of the model parameters.

\begin{figure}[!ht]
    \centering
    \includegraphics[width=0.95\textwidth]{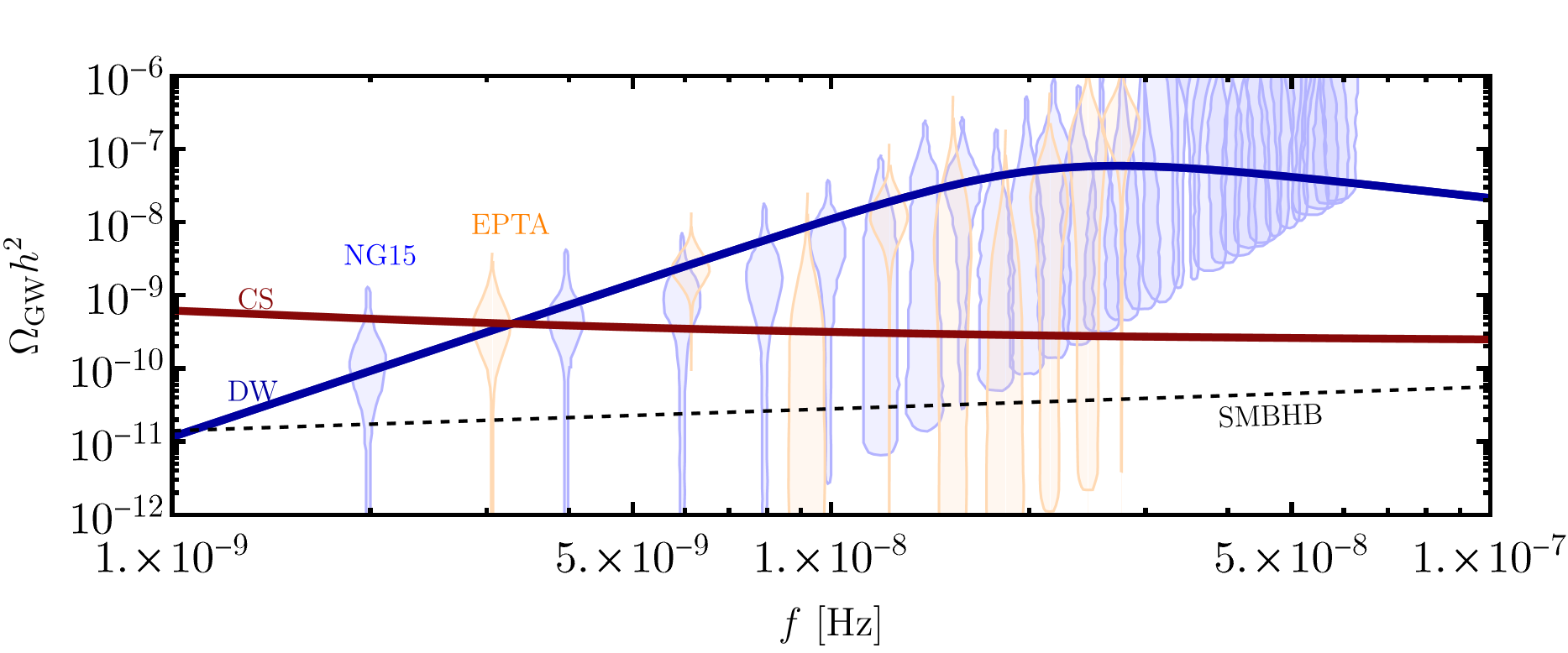}
    \caption{Gravitational wave spectrum at PTA frequencies from the best fit parameters of the Dirac mass model (benchmark point \Circled{1} in Table~\ref{table:BP}) labeled DW, and Majorana mass model ($B-L$ breaking scale $\Lambda = 1.2 \times 10^{14}$~GeV) labeled CS. Recent results from NANOGrav 15yr data (blue violins) and EPTA data (orange violins) are also shown. The dashed line is the signal from SMBHBs with $A_{\rm BHB} = 10^{-15.6}$ and $\gamma_{\rm BHB} = 4.7$~\cite{NANOGrav:2023hvm}.}
    \label{fig:GWDWNG15}
\end{figure}

For the Majorana seesaw model, we show the $68\%$ (darker blue) and $95\%$ (lighter blue) Bayesian credible regions for the single parameter $\Lambda$. For the Dirac seesaw model, we show both the reconstructed 1D marginalized distributions (diagonal plots) and the 2D distribution (off-diagonal), along with $68\%$ (darker blue) and $95\%$ (lighter blue) Bayesian credible regions. In the same plots, we also show the fit to the combined contribution (red) from the respective models and from inspiraling supermassive black hole binaries (SMBHBs)~\cite{NANOGrav:2023hvm}
\begin{align}
    \Omega_{\rm GW} h^2 \approx \frac{2\pi^2 h^2 A_{\rm BHB}^2 f^5}{3H_0^2} \left(\frac{f}{\rm{yr}^{-1}}\right)^{-\gamma_{\rm BHB}} {\rm yr}^3\,, \label{SMBHBgw}
\end{align}
where $H_0 = h \times 100\ \text{kms}^{-1}\text{Mpc}^{-1}$ is the Hubble constant, using the best fit parameters from numerical simulations $A_{\rm BHB} \simeq 10^{-15.6}$ and $\gamma_{\rm BHB} \simeq 4.7$~\cite{NANOGrav:2023hfp}. We notice that the inclusion of the SMBHB contribution does not noticeably enlarge the parameter space of either model. For the Majorana seesaw model, the $U(1)_{B-L}$ breaking scale $\Lambda$ is in the range $10^{14}$~GeV-$10^{14.16}$~GeV at the $68\%$ credible level. For the Dirac seesaw model, the $Z_2$ breaking scale $u$ and bias potential $V_{\rm bias}$ are between $10^{5.5}$~GeV-$10^{-5.76}$ GeV and $10^{-2.8 }\ \text{GeV}^4$-$10^{-1.8 }\ \text{GeV}^4$, respectively, at the $68\%$ credible level.

We choose the best fit points of the respective models and plot the GW spectrum in Fig.~\ref{fig:GWDWNG15}.
The Majorana seesaw model with $B-L$ symmetry broken at $\Lambda = 1.2 \times 10^{14}$ GeV generates the maroon curve, and the Dirac seesaw model with benchmark point \Circled{1} yields the dark blue curve at PTA frequencies. The blue violins show the NANOGrav 15yr data and the orange violins show the second data release from EPTA~\cite{EPTA:2023sfo, EPTA:2023akd, EPTA:2023fyk, EPTA:2023gyr, EPTA:2023xxk, EPTA:2023xiy}. For comparison, we also show the signal (black dashed line) from SMBHBs with the best fit values of $A_{\rm BHB}$ and $\gamma_{\rm BHB}$. The spectral slope of the cosmic string signal from the Majorana mass model provides an even poorer fit to the PTA signal than the SMBHB signal. On the other hand, the domain wall signal from the Dirac mass model fits the PTA result remarkably well. This suggests that Majorana mass generation from the spontaneous breaking of a gauged $B-L$ symmetry at around $10^{14}$~GeV is not favored by PTA data, while Dirac mass generation from the spontaneous breaking of the $Z_2$ symmetry in Eq.~\eqref{eq:Lagrangian} at around $10^{5-6}$~GeV is favored. While this statement relies on several assumptions about the astrophysical background in PTA data, and is not necessarily an \emph{inevitable} outcome of Dirac or Majorana neutrino mass generation, it provides an example of the complementary information one may expect from GW signatures of the respective seesaw models. In the absence of any direct evidence of the nature of neutrino mass, such indirect probes are important and encourage further exploration in this direction.

\section{Discussion and outlook \label{sec:conclusion}}
We have considered a novel possibility to distinguish between Dirac and Majorana neutrino mass through the difference in the gravitational wave spectra, assuming a seesaw mechanism in each case. 
For definiteness, we considered two simple classes of the respective type-I seesaw models, both inspired by the assumption that small neutrino masses be generated without tiny Yukawa couplings. The Majorana seesaw is assumed to involve spontaneous breaking of lepton number, generating cosmic strings, while the Dirac seesaw is assumed to involve spontaneous breaking of a discrete $Z_2$ symmetry (the minimal choice) leading to domain walls. 
The resulting very different shapes of the GW spectra, after the cosmological relics decay,
flat in the Majorana seesaw case, and peaked in the Dirac seesaw case may help distinguish between the two seesaw mechanisms. While the method by itself is not conclusive, it may provide valuable additional information about the mass generation mechanism of neutrinos, which is combinable with other indirect evidence to determine the nature of neutrino mass.

We emphasize that the primary difference between Majorana and Dirac neutrino masses is whether or not lepton number is broken. In the case of Majorana masses, it is broken by two units. In a well motivated scenario, the lepton number symmetry, or equivalently, the $B-L$ symmetry, is exact at ultraviolet scales, but is spontaneously broken when a scalar charged under the gauged $U(1)_{B-L}$ symmetry gets a nonzero VEV. This gives Majorana mass to the right-handed neutrinos, which further generates small masses for the SM neutrinos via the type-I seesaw mechanism. The breaking of the $U(1)_{B-L}$ symmetry triggers the creation of cosmic strings in the early Universe. The string network loses energy via the production of string loops, some of which emit gravitational waves. The GWs have a flat spectrum over a wide range of frequencies, and their amplitude is related to the scale of symmetry breaking. Hence, the detection of a flat spectrum of stochastic gravitational wave background may imply a Majorana nature of the neutrinos and shed light on the scale at which such masses are generated. 

On the other hand, the Dirac seesaw mechanisms to generate a small Dirac mass for the SM neutrinos utilize the effective operator $\bar{L} \nu_R H \sigma$. To keep lepton number symmetry unbroken, and to prohibit a tree-level Dirac mass term for the SM neutrinos requires  $\nu_R$ and $\sigma$ to be non-trivially charged under a $Z_2$ symmetry, which is spontaneously broken by the VEV of $\sigma$. The breaking of a discrete symmetry creates a domain wall network, which poses a threat to the standard cosmology if it is long-lived and/or dominates the energy density of Universe. Domain walls can be made to annihilate by softly breaking the discrete symmetry, thereby lifting the degeneracy between the $Z_2$ symmetric vacua and creating a bias potential that tends to collapse the walls. This leads to a characteristic GW signal peaked at a frequency determined by the scale of spontaneous and soft symmetry breaking. Interestingly, there are several ultraviolet realizations of the Dirac seesaw operator at the tree-level and at one-loop level. The characteristics of the GW signals produced in the realizations of the Dirac seesaw operator are identical, since these are solely determined by the potential of the scalar $\sigma$, as long as $\sigma$ does not mix with other scalars. We have shown that depending on the parameter space, such signals may be probed by various terrestrial and satellite-based interferometers. We compared the predictions of the two classes of mass generation models to the NANOGrav dataset and found that Dirac seesaw models are favored.

The scope of our proposal is limited in various respects. While our setup is well-motivated, it is possible that the $B-L$ symmetry is explicitly broken in the case of Majorana neutrinos, or that the scale of breaking is small enough that the signal is undetectable. Since our discussion is focused on a subclass of Dirac and Majorana mass generation models, our results are representative of the type of models, and not of the specific nature of the neutrinos (which can be probed by neutrinoless double beta decay experiments). It is also arguable that other models, unrelated to neutrino mass generation, can produce similar signals from cosmic strings or domain walls.  Furthermore, we have assumed a typical radiation-dominated Universe after inflation, and modifications to the standard cosmic evolution may affect the GW spectra. Nevertheless, the model classes we have investigated are simple, minimalistic, and highly predictive under the set of assumptions made. 
Hence, our work should be viewed as an attempt to extract whatever meaningful information one can from GW signals in the absence of a positive result from neutrinoless double beta decay experiments. In particular, given that a nonobservation of neutrinoless double beta decay does not necessarily imply that neutrinos are Dirac particles, the peaked domain wall signal may provide indirect evidence of the Dirac nature of neutrinos when combined with data from other terrestrial experiments. 


\acknowledgments
{We thank Anish Ghoshal for a helpful discussion, and Xuce Niu and Andrea Mitridate for assistance with \texttt{PTArcade}. S.F.K. would like to thank IFIC, University of Valencia for hospitality.
S.F.K. and M.H.R. acknowledge  support from the STFC Consolidated Grant ST/T000775/1, and from the European Union's Horizon 2020 Research and Innovation
Programme under Marie Sklodowska-Curie grant agreement HIDDeN European
ITN project (H2020-MSCA-ITN-2019//860881-HIDDeN). M.H.R. is also supported by the Spanish AEI-MICINN PID2020-113334GB-I00/AEI/10.13039/501100011033 and Generalitat Valenciana project CIPROM/2021/054. D.M. is supported in
part by the U.S. Department of Energy under Grant No. DE-SC0010504. 
}

\bibliography{references}
\newpage
\bibliographystyle{JHEP}
\end{document}